%% file: preprint040827.tex
\documentclass{aa}
\usepackage{graphicx}
\usepackage{txfonts}
\usepackage{natbib}
\newcommand{\Ks}{K_\mathrm{s}}
\begin{document} 
   \title{XMM-Newton and VLT observations of the afterglow of GRB\,040827\thanks{Based on observations with XMM-Newton,
 an ESA science mission with instruments and contributions directly  funded by ESA member states and the USA (NASA),
 and on observations collected at the European Southern Observatory, Paranal, Chile, under proposal 073.D-0255.}}

%   \subtitle{I. Overviewing the $\kappa$-mechanism}

\author{A.~De Luca\inst{1}
\and
A.~Melandri\inst{2}
\and
P.A.~Caraveo\inst{1}
\and
D.~G\"otz\inst{1}
\and
S.~Mereghetti\inst{1}
\and
A.~Tiengo\inst{1}
\and
L.A.~Antonelli\inst{2}
\and
S.~Campana\inst{3}
\and
G.~Chincarini\inst{3}
\and
S.~Covino\inst{3}
\and
P.~D'Avanzo\inst{3,4}
\and
A.~Fernandez-Soto\inst{5}
\and
D.~Fugazza\inst{3}
\and
D.~Malesani\inst{6}
\and
L.~Stella\inst{2}
\and
G.~Tagliaferri\inst{3} 
%\and
%et al.\inst{3}
          }

   \offprints{A. De Luca, deluca@mi.iasf.cnr.it}

   \institute{INAF - Istituto di Astrofisica spaziale e Fisica Cosmica, sez. di Milano ``G.Occhialini'',
Via Bassini 15, I-20133, Milano, Italy 
\and
INAF - Osservatorio Astronomico di Roma, Via di Frascati 33, I-00040 Monteporzio Catone (Rome), Italy. 
         \and
INAF - Osservatorio Astronomico di Brera, Via Bianchi 46, I-23807 Merate (LC), Italy 
\and
Universit\`a degli Studi dell'Insubria, Dipartimento di Fisica e Matematica, Via Valleggio 11, I-22100 Como, Italy
\and
Observatori Astr\`onomic, Universitat de Val\`encia, E-46100, Spain
\and
International School for Advanced Studies (SISSA-ISAS), via Beirut 2-4, I-34014 Trieste, Italy
             }
   \date{Received; accepted}

   \abstract{
The field of the Gamma-Ray Burst GRB\,040827 was observed with XMM-Newton and with the ESO/VLT
starting $\sim6$ and $\sim12$ hours after the burst, respectively. A fading X-ray
%(F$\propto t^{-\delta}$, $\delta=1.36\pm0.1$)
afterglow is clearly detected with the XMM-Newton/EPIC instrument,
with a time decay $t^{-\delta}$, with $\delta=1.41\pm0.10$. Its spectrum is well described by a power law
(photon index $\Gamma=2.3\pm0.1$) affected by an absorption largely exceeding (by a factor $\sim$5) the expected
Galactic one, 
%in the burst direction, 
requiring the contribution of an intrinsic, redshifted absorber. 
%to be added to the Galactic N$_H$.  
In the optical/NIR range, the afterglow emission was observed in the $K_{\rm s}$ band, as
a weak source superimposed to the host galaxy, with magnitude $K_{\rm s} =
19.44 \pm 0.13$ (12 hours after the GRB, contribution from the host subtracted);
in other bands the flux is dominated by the host galaxy. 
Coupling constraints derived from X-ray spectral fitting
and from photometry 
of the host, we estimated a gas column density in the range
(0.4-2.6)$\times10^{22}$ cm$^{-2}$
in the GRB host galaxy, likely located at a redshift $0.5<z<1.7$.
GRB\,040827 stands out as the best example of an X-ray afterglow with intrinsic absorption.

   }

\titlerunning{GRB\,040827}

   \maketitle
%
%________________________________________________________________

\section{Introduction}
The discovery of afterglow emission at X-ray, optical and radio
wavelengths opened a new era in the study of the mysterious 
Gamma-Ray Burst (GRB) phenomenon. Afterglows' observations yielded most of our 
current understanding of long ($>$2 s) GRBs, from their energetics
to the evidence of their association with the death of massive stars 
\citep[see, e.g.][for comprehensive reviews]{vanparadijs00,hurley03,piran04}. 
A fast and accurate localization of GRBs is required for successful follow-up
studies of their afterglows. After the end of the BeppoSAX mission, the INTEGRAL satellite proved
to be very powerful in such task, thanks to the INTEGRAL Burst Alert System \citep[IBAS,][]{mereghetti03b},
providing near real-time, precise ($\sim2'-3'$) positioning of about one burst
per month. This allows for follow-up observations of GRBs with the most sensitive
facilities at different wavelengths. Here we report 
on the X-ray and  optical/NIR
studies of the afterglow of GRB 040827 as seen by the XMM-Newton 
observatory and the ESO very Large Telescope (VLT), respectively.
\section{GRB\,040827}
GRB\,040827 was discovered by the INTEGRAL Burst Alert System 
on 2004, August 27 at 11:50:48 
UT.
The burst, detected 
with the IBIS/ISGRI instrument in the 15-200 keV band at $\alpha=$15$^{\rm h}$16$^{\rm m}$59.8$^{\rm s}$, 
$\delta=$-16$^\circ$08$'$21$''$, with 
an uncertainty of 2.5\arcmin, 
had a duration of 40\,s \citep{Mere04c,Mere04d}. 
The 
peak flux in the 20-200 keV 
band was estimated to be 0.6 photons\,cm$^{-2}$\,s$^{-1}$, corresponding to $6.0 \times 10^{-8}$ 
erg\,cm$^{-2}$\,s$^{-1}$ \citep{GotzMer04}.

An XMM-Newton observation of the field of
GRB\,040827 started on August 27 at 
18:07:56 UT, 6 h 18 min after the burst.
Soon after the beginning of 
the observation, analysis of real-time 
data showed an X-ray source 
\citep[XMMU J151701.3-160828,][]{RodrJua04} within the
INTEGRAL error circle, with an offset of $\sim$1 arcmin
with respect to the nominal aimpoint\footnote{See 
http://xmm.vilspa.esa.es/external/xmm\_news/items/grb040827/ index.shtml}.
The telescope was
therefore re-pointed $\sim$4300 s after the
beginning of the observation. The overall observing time 
was of 53.2 ks.
The source XMMU J151701.3-160828 was found to be variable
\citep{rodrgonz04}, 
and therefore likely associated to the afterglow of GRB\,040827. 
Such an identification was 
strenghtened by the discovery of a fading NIR transient (see below).

Optical imaging of the INTEGRAL error box started immediately after 
the IBAS alert.
\citet{GladBerg04} and
\citet{Tanv04a} reported the presence inside the XMM-Newton error circle
of a faint, extended source
with $K_{\rm s} \sim 19.4$, named XMM\,2, which was subsequently found to be
variable \citep{KaplBerg04,Tanv04b,Males04c}.
No radio emission was detected from the source, with a $2\sigma$ upper limit
of 70~$\mu$Jy, 4.5 days after the GRB \citep{soderberg04}.

A preliminary report of the 
XMM-Newton
observation was given by \citet{DeL04b},
but the results were affected by problems in the data possibly due to 
the satellite re-pointing. An improved attitude solution was then made 
available by the XMM-Newton SOC \citep{schartel04}.
We present here 
a detailed study of the 
X-ray
dataset, coupled to an 
analysis of the complete set of observations performed  with the VLT.

\section{XMM-Newton data analysis and results}
\label{xmm}

We have analyzed data collected by the 
European Photon Imaging Camera (EPIC). It 
consists of a PN detector \citep{struder01}, with
a collecting area of $\sim900$ cm$^2$ at 1.5 keV,
and of 2 MOS detectors \citep{turner01}, with a collecting
area of $\sim450$ cm$^2$ each at 1.5 keV.
All the cameras were operated in their full frame mode
(providing imaging across the full 15\arcmin radius field of view,
with a time resolution of 73 ms in the PN and 2.6 s in the MOS),
with the thin optical filter.

The Observation Data Files (ODFs) were retrieved from
the XMM-Newton Science Archive and were processed with
the XMM-Newton Science Analysis
Software (XMM-SASv6.0.0), using standard pipeline tasks
({\em epchain} and {\em emchain} for the PN and the MOS,
respectively) to produce calibrated photon lists, with 
reconstructed energy, time of arrival and coordinates on the field of view for each 
detected event.

The X-ray afterglow of GRB\,040827
is clearly detected in all the cameras. 
Cross-correlating several field sources' positions with the USNO-B1 catalog
improved the EPIC astrometry and yielded a source position at $\alpha=$15$^{\rm h}$17$^{\rm m}$01\fs46,
$\delta=$-16\degr08\arcmin27\farcs2, with a 1$\sigma$ uncertainty of 1\farcs5.
The position is fully consistent with that
of the fading NIR afterglow. 
%(RA=15h 17m 01.35s, Dec=-16d 08$'$ 28.6$''$).

The surface brightness radial profile of the source, computed 
for each EPIC detector, was fitted with a King function.
The best fitting parameters were found to be consistent with 
the instrumental Point Spread Function \citep{read04}.
A claim of possible diffuse emission based on quick look data analysis
\citep{RodrJua04}
was 
clearly ascribable
to problems in
the events' attitude correction.

Source photons were extracted 
from a circle of 35 arcsec radius, containing $\gtrsim$85\% of the events.
Background events 
were selected from source-free regions 
of
the same CCD chip where the target is imaged, 
following standard prescriptions from the calibration team \citep{kirsch04}:
for the PN we used a circle of
50 arcsec radius, located at the same distance from the readout node 
as the source region; for the MOS we used an annulus centered on the source
position with inner and outer radii of 90 and 200 arcsec.

%----------------------------------------------------------- 
   \begin{figure}
   \centering
   \includegraphics[height=8cm,angle=-90]{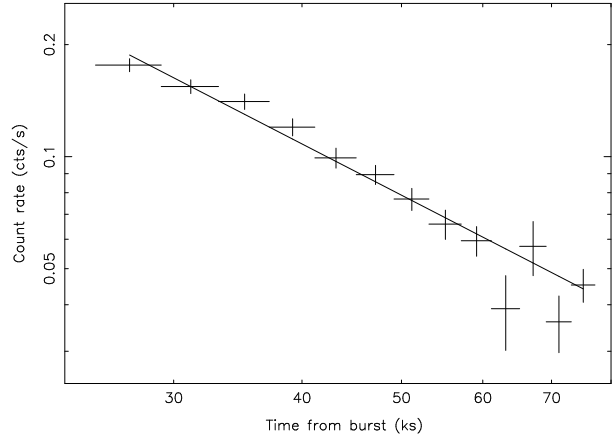}
      \caption{Background-subtracted EPIC light curve (0.2-10 keV) of the afterglow of GRB\,040827.
Data have been fitted with a power law decay 
$F\propto~t^{-\delta}$, with a best fit value of $\delta=1.41\pm0.10$
($\chi^2_{\nu}=1.45$, 11 d.o.f.). }
         \label{lc}
   \end{figure}
%
%______________________________________________________________

The source is clearly seen to fade during the observation.
The background-subtracted light curve extracted from combined MOS and PN data is shown 
in Fig.~\ref{lc}. The time evolution of the observed count rate 
is consistent with a power law decay, $F\,\propto\,t^{-\delta}$, with $\delta$=1.41$\pm$0.10
($\chi^{2}_{\nu}=1.45$, 11 d.o.f.). A somewhat better description ($\chi^{2}_{\nu}=0.79$, 9 d.o.f.)
of the decay is obtained with a 
broken power law model, with $\delta_1=0.9^{+0.3}_{-0.5}$, $\delta_2=1.67^{+0.25}_{-0.20}$ and
break time $t_{b}=36\pm6$ ks after the burst. With a  simple F-test we evaluated the fit
improvement to be significant at $\sim97.3\%$ level (well below $3\sigma$).

In order to perform the spectral analysis, we screened data from 
time intervals affected by high particle background episodes
(soft proton flares),
obtaining
good exposure times of 30.0 ks,
38.8 ks and 37.3 ks
for the PN, MOS1 and MOS2 cameras, respectively. Spectra for both source and
background were extracted from the regions described above. Source spectra
were rebinned in order to have at least 25 counts per channel.
Ad-hoc response matrices and effective area files were computed with the
SAS tasks {\em rmfgen} and {\em arfgen}\footnote{ 
Possibly owing to the $\sim1'$ change in pointing direction during the observation,
to generate the effective area file it was necessary to describe the source
region (in detector coordinates) using a larger number of bins than in standard cases
(namely, {\em arfgen} parameters {detxbins} and detybins were set to 30 wrt. a default value of 5).}.

The spectral analysis was performed
using XSPEC v11.3. Spectra from the three detectors were fitted symultaneously in
the range 0.2-9 keV.
All the errors on spectral parameters 
reported in this section are at 90\% confidence level for a single interesting parameter.

An absorbed power law model (i.e. $dN/dE=e^{-\sigma(E)N_{\rm H}}\,K(E/E_0)^{\Gamma}$, where the photoelectric
cross sections $\sigma(E)$ are from \citet{balucinska92}, with updated He cross-sections, and $E_0=1$ keV) 
yields a reduced
$\chi^{2}_{\nu}$=1.35 (145 dof), with 
photon index
$\Gamma=2.4\pm$0.1 and  column density $N_{\rm H} = (3.7 \pm 0.2) \times10^{21}~{\rm cm}^{-2}$.
Since the expected Galactic absorption in this direction varies from
$\sim5\times10^{20}~{\rm cm}^{-2}$ \citep{schlegel98} to
$\sim8\times10^{20}~{\rm cm}^{-2}$
\citep{dickey90}, we conclude that our best fitting $N_{\rm H}$ value is significantly higher than the 
Galactic one.

This is confirmed by the study of
the brightest serendipitous source in the field,
located at $\sim7'$ from the afterglow.  
Its spectrum is well fitted by an absorbed power law with 
$\Gamma$=2.2$\pm$0.2 and 
$N_{\rm H}=(9\pm3)\times10^{20}~{\rm cm}^{-2}$.
The position of such source was cross-correlated with multiwavelength 
catalogs, 
yielding
(within $\sim$1$''$) a USNO-B1 
source with a magnitude  $R \sim19.3$, corresponding to a ratio
$F_{\rm X} / F_{\rm opt} \sim 8$, in the typical range of active galactic nuclei \citep{krautter99}.
Moreover, a coincident radio source (NVSS J151717-160242) was found in the NED database.
We conclude that such source is very likely an AGN and that the observed X-ray 
absorption 
is a reliable and independent estimate
of the Galactic $N_{\rm H}$ in this direction.

On the other hand,
 the Galactic value of the column density
is totally uncompatible with the afterglow spectrum: forcing the $N_{\rm H}$ value to 
$8\times10^{20}~{\rm cm}^{-2}$ 
we obtain
$\chi^{2}_{\nu}$ of 3.86 (146 dof).

%----------------------------------------------------------- 
   \begin{figure}
   \centering
   \includegraphics[height=8cm,angle=-90]{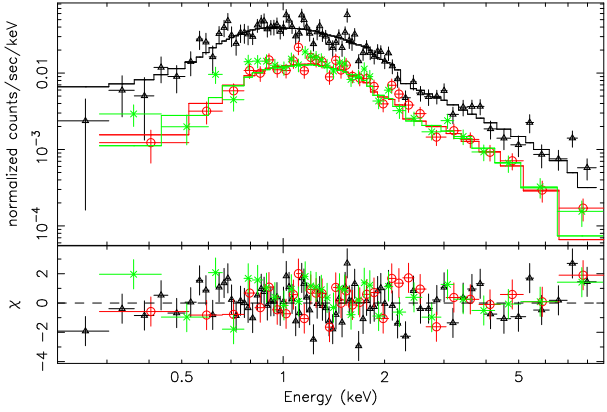}
      \caption{EPIC observed spectra plotted together with the best fitting model convolved with
the instrumental responses (upper panel); residuals in units of statistical 
standard errors (lower panel). 
Triangles, circles and crosses represent data from the PN, MOS1 and MOS2 
cameras, respectively.
}
         \label{spectra}
   \end{figure}
%
%______________________________________________________________

The best description of the afterglow data 
was obtained by 
%fixing $N_H$ 
adding
to the Galactic 
%value
$N_{\rm H}$
($8\times10^{20}~{\rm cm}^{-2}$) 
a redshifted neutral absorption component, $N_{\rm H,z}$. 
In this case the spectral shape is described as 
$dN/dE=e^{-\sigma(E)N_{\rm H}}\,e^{-\sigma[E(1+z)]N_{\rm H,z}}\,K(E/E_0)^{\Gamma}$,
where $N_{\rm H}$ is fixed to the Galactic value of $8\times10^{20}~{\rm cm}^{-2}$
and $N_{\rm H,z}$ represents the gas column density in the host.
In the redshift interval $0<z<3$, 
we obtained 
the best fit ($\chi^{2}_{\nu}$ of 1.18 for 144 dof - see Fig.~\ref{spectra}) 
for an
intrinsic absorbing column
$N_{\rm H,z}\sim1.0\times10^{22}$ cm$^{-2}$ 
at
a redshift $z \sim 0.9$, 
and a power law photon index $\Gamma = 2.3 \pm 0.1$.
The observed 0.2-10 keV flux is $\sim2.2\times10^{-13}$ erg cm$^{-2}$
s$^{-1}$; the corresponding unabsorbed flux is $\sim4.9\times10^{-13}$ erg
cm$^{-2}$ s$^{-1}$.

\begin{figure}
   \centering
   \includegraphics[height=8cm,angle=-90]{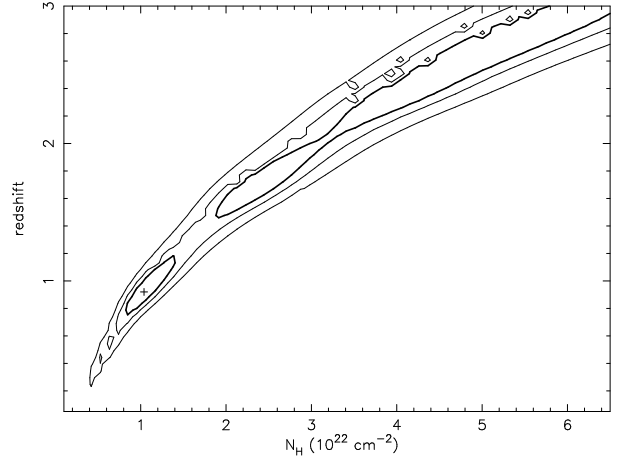}
      \caption{Confidence contours (68\%, 90\% and 99\%) 
for the redshift and the column density of the 
redshifted cold absorber derived from the fit to EPIC spectra. 
}
         \label{confcont}
   \end{figure}

Both  
$N_{\rm H,z}$ and $z$ are not well constrained
towards large values, owing to their strong correlation.
The contour plot for the redshifted absorber column density $N_{\rm H,z}$ vs. 
the redshift $z$, shown in Fig.~\ref{confcont}, 
allows us to estimate that $N_{\rm H,z}>4\times10^{21}$ cm$^{-2}$ and $z>0.5$
at 90\% confidence level (for two
parameters of interest).

%----------------------------------------------------------- 
%
%______________________________________________________________

To evaluate the statistical significance of the additional  $N_{\rm H,z}$,
we used the F-test, comparing the 
models 
with: 
(i) $N_{\rm H}$ fixed
to the Galactic value; (ii) $N_{\rm H}$ allowed to vary; (iii)  
redshifted absorber added to the fixed Galactic $N_{\rm H}$. 
Results are reported in Table~\ref{ftest}. The high significance of 
the intrinsic absorption is evident even in the pessimistic assumption
of a Galactic $N_{\rm H}$ 50\% higher than expected, a value which
could be marginally consistent with the absorption seen in the spectrum 
of the serendipitous AGN.

\begin{table}
\centering\caption{\label{ftest} Significance of additional $N_{\rm H}$. The F-test was used
to compare different models for the X-ray absorption: $N_{\rm H}$ fixed
to the Galactic value (`Gal. $N_{\rm H}$'); $N_{\rm H}$ allowed to vary (`free $N_{\rm H}$'); 
redshifted absorber added to the fixed Galactic $N_{\rm H}$ (`$N_{\rm H,z}$'). The value of F and the probability 
of a chance occurrence of the fit improvement are reported. The high significance of the extra absorption 
is apparent also in the pessimistic assumption of an actual Galactic $N_{\rm H}$ 50\% higher than expected following
\citet{dickey90}.}
\begin{tabular}{ccc}
\hline 
Absorption models & F value & $P$(F)  \\ \hline
Gal. $N_{\rm H}$ vs. free $N_{\rm H}$ & 274 & $3.0\times10^{-35}$ \\
Gal. $N_{\rm H}$ vs. $N_{\rm H,z}$ & 167 & $2.8\times10^{-38}$ \\ \hline
Gal. $N_{\rm H}$+50\% vs. free $N_{\rm H}$ & 187 & $6.8\times10^{-28}$ \\
Gal. $N_{\rm H}$+50\% vs. $N_{\rm H,z}$ & 118 & $5.3\times10^{-31}$ \\ \hline
Free $N_{\rm H}$ vs. $N_{\rm H,z}$ & 21.5 & $8.0\times10^{-6}$ \\
\end{tabular}
\end{table}

We note that the residuals plotted in Fig~\ref{spectra} (lower panel) 
show several small wiggles.
In order to search for possible emission lines in the afterglow's spectrum, 
we tried to add gaussian lines, both single and multiple (up to 5 at a time).
The width was fixed (smaller than the instrumental energy resolution);
the central energy, together with the line flux, was allowed to vary in the range 0.5-5 keV, where
most of the statistics is collected.
No significant lines were detected
in such range in the combined MOS/PN dataset.
The upper limits ($3\sigma$) on the equivalent width of any emission line are
$\sim$60 eV and  $\sim$250 eV 
in the ranges 0.5-2 and 2-5 keV, respectively.

Thermal models (requiring, in any case, extra absorption wrt. the Galactic $N_H$) 
cannot fit the data equally well.
%give a worse fit to the spectra.
A redshifted, optically thin
plasma model (MEKAL in XSPEC), with $z$ linked to the redshift of the neutral
cold absorber $N_{\rm H,z}$, yields rather poor  results ($\chi^{2}_{\nu}$ of 1.89 for 144 dof)
when abundances are fixed to solar system values. If abundances are allowed to vary, 
better results can be obtained. However, in any case the resulting $\chi^{2}_{\nu}$ is worse ($>$1.4) than
for the simple power law model.

We tried also a composite model 
encompassing a power law 
as well as a plasma emission model, 
but the resulting $\chi^{2}$ is similar to that obtained using the simpler power law model.

We searched for a possible spectral evolution,
using the power law plus redshifted absorber model.
We divided the 
observation in two time intervals of $\sim13$ and $\sim22$ ks, 
in order to have approximately the same number of source photons in each subset,
and extracted the corresponding spectra for each EPIC detector.
In a symultaneous fit, a simple variation of the power law intensity
yields a good description of the data ($\chi^{2}_{\nu}$ of 1.0 for 180 dof).
No significant evolution is seen for the absorbing column $N_{\rm H,z}$ and the photon index.
No significant deviations from the smooth
continuum are seen in the time-resolved data.

\section{VLT observations}
\label{vlt}
\begin{figure}\centering
  \includegraphics[width=\columnwidth,keepaspectratio]{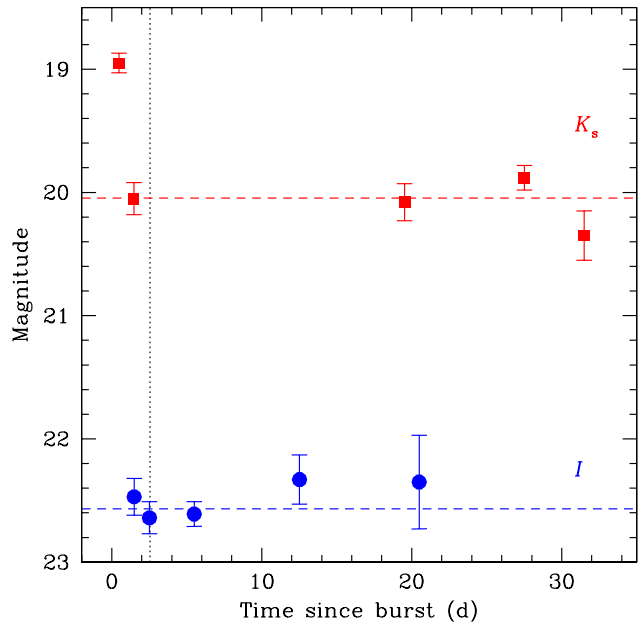}
  \caption{Light curve of the optical counterpart of
  GRB\,040827. Squares and circles correspond to $K_{\rm s}$- and $I$-band
  measurements respectively. The first $K_{\rm s}$-band point is the average of our 
first two exposures. The horizontal dashed lines show the host
  galaxy contribution, and the vertical dotted line marks the epoch of
  our spectrum.\label{fg:optlc}}
\end{figure}

\input{optlognew.tex}

Near-infrared (NIR) observations of the error box of GRB\,040827 started
as soon as possible after the notification of the trigger. We used the
ISAAC camera installed at the Nasmyth focus of the ESO VLT--UT1 (Antu) located at Paranal
(Chile). Since the field of view of ISAAC ($2.5\arcmin \times
2.5\arcmin$) is not large enough to image the whole INTEGRAL error box,
a mosaic of four exposures was obtained.
In the following nights, further optical and NIR observations
were secured (Tab.~\ref{tb:optlog}). Optical data were acquired with the
FORS\,1 and FORS\,2 instruments mounted at the Cassegrain focus of the VLT--UT2 and UT1
respectively. Data reduction was performed using the standard
procedures. Both aperture and profile photometry \citep[using DAOPHOT -][]{stetson87} 
were adopted to extract instrumental
magnitudes. Absolute calibration was achieved by observing standard
fields (in the optical) and by using the 2MASS catalog (in the
infrared). The consistency of the zeropoint was checked in different
nights. In some cases the photometric errors are larger than usual,
since the object could be observed just for a small part of the night,
at a large airmass.

The source reported by  \citet{GladBerg04}
and  \citet{Tanv04a}
was clearly detected in our images at the coordinates $\alpha = 
15^{\rm h}17^{\rm m}01\fs34$, $\delta = -16\degr08\arcmin29\farcs1$ 
(J2000, $\sim 0.2\arcsec$ uncertainty, based on 9 nonsaturated USNO 
stars). Its magnitude was $K_{\rm s} = 18.95 \pm 0.08$ on Aug 27.98 
(average of our first two measurements), and it showed a dimming by 
$1.10 \pm 0.15$~mag in the $K_{\rm s}$ filter, between 2004 Aug 27.98 
and 28.98 (0.49 and 1.48 days after the GRB, respectively), thus 
confirming the reported variation. Since the seeing was similar in the 
two epochs ($\approx 0.9\arcsec$), this result is robust even if in our 
best-seeing images there are hints that the object is slightly extended 
(with a size of $\sim 0.5\arcsec$).

Long-term monitoring was performed in the $I$ and $K_{\rm s}$ bands (see
Fig.~\ref{fg:optlc}). No significant variation of the source was seen
from $1.5$ up to $\sim 30$~days after the burst, indicating that the
source was dominated by the host galaxy. This is consistent with the
source being extended. In particular, there is no sign of afterglow
emission in the $R$ and $I$ bands (for which we have not 
data during the first night). 
In our
early $J$ and $H$ images, a contribution from the afterglow is likely
present, but the host galaxy is contributing by a significant, unknown
amount, hampering the determination of the intrinsic afterglow brightness.
After averaging over the available measurements, we get for the
host galaxy $R = 23.10 \pm 0.12$, $I = 22.57 \pm 0.06$ and $K_{\rm s} = 20.05
\pm 0.07$. Therefore, we can estimate the afterglow magnitude to be $K_{\rm s}
= 19.44 \pm 0.13$ and $K_{\rm s} > 21.02$ (3$\sigma$) on 2004 Aug 27.98 and
2004 Aug 28.97  respectively. Assuming a power law decay ($F(t) \propto
t^{-\delta}$), the decay index is constrained to be $\delta > 1.30$
(3$\sigma$). 
We caution that this value is computed from just two measurements,
and may be influenced by local deviations from the regular decay and/or
measurement errors.
This limit is consistent with the X-ray decay
index.

A spectrum of the host galaxy was acquired with FORS\,1,
starting on 2004 Aug 30.00 (2 hours exposure with the grism
300V). The observing conditions were not good, since the object
was visible only at relatively large airmass ($\sim 2$) and with
full (100\%) Moon.
The object was marginally detected in the range $4600 \div 8600$~\AA,
characterized by a very weak continuum with no distinguishable
features. Therefore, no redshift determination was possible. 
The 2-$\sigma$
upper limit for any emission line is $10^{-16}$~erg~cm$^{-2}$~s$^{-1}$.

Costraints on the distance of the galaxy can be put using the
photometric information. With just two colors, photometric redshift
techniques are not very effective. The algorithm developed by
\citet{fernandez99}
 yields just loose
constraints $1.0 < z < 3.5$, with an additional permitted range $0.1 < z
< 0.3$ (where the fit is  worse).

However, we can use the apparent luminosity to get a rough estimate of
the distance. To this extent, we have choosen several catalogs of galaxies
providing for each object both the photometry and the redshift. From these
data, we computed the redshift distribution of the sources which have
a magnitude comparable to that of the host galaxy of GRB\,040827. Data
were collected from the VIMOS VLT deep survey \citep[VVDS;][]{lefevre04}, 
the Hubble deep field \citep{fernandez99}, the
FORS deep field \citep{heidt03,noll04} and the Caltech
faint galaxy redshift survey \citep{cohen00}. We found that in
the $\Ks$ band the distribution is quite broad (peaking at $z = 0.9$),
and it doesn't allow to put significant contraints. Tighter limits can
be set by looking at the $I$-band apparent luminosity; the distribution
of 127 galaxies from the VVDS within 0.15~mag from the GRB host is well
represented by a Gaussian centered at $z = 0.77$ with a dispersion of
$0.31$. The VVDS is particularly suited for this kind of work,
since it has a high completeness in redshift determination, and because
it succesfully crosses the so-called redshift desert $1.5 < z < 2.7$
\citep{lefevre04}, where redshift determination is difficult due
to the lack of clear features. Thus, based on a statistical analysis,
we may put a 3$\sigma$ upper limit $z < 1.7$.  The same procedure in
the $R$ band provides an even narrower range ($z = 0.66 \pm 0.23$),
however based on a smaller sample. We caution that our object,
since it hosted a GRB, may not be a typical galaxy, therefore our
indicator may be biased. It is anyway reassuring that the estimate
from both the $R$ and $I$ bands are consistent. Thus, the different
indications coming from both optical and X-ray data favor a redshift not
far from unity. This is consistent with the redshift distribution of GRBs
\citep[e.g.][]{bloom03}.

Within the photometric errors, no late-time rebrightening is apparent in
our data. However, even a bright supernova like SN\,2003lw \citep{malesani04}
 peaked at $J \sim 18.8$ (at $z = 0.1055$), which would translate
to $K_{\rm s} \sim 22.8$ at $z \sim 1$ (the observed $K_{\rm s}$ band roughly
corresponds to the $J$ at this redshift). Such a SN would be $\sim 13$
times fainter than the host galaxy, hence quite difficult to detect.

\section{Discussion}
XMM-Newton/EPIC data show unambiguously that 
the spectrum of the X-ray afterglow is 
affected by an absorption significantly higher than the Galactic value.

As shown in Sect.~\ref{xmm}, assuming solar system abundances,
the redshifted absorber model allows to estimate that at 90\% confidence the 
local column density in the GRB host $N_{\rm H,z}$ is higher than $4\times10^{21}$ cm$^{-2}$ 
and its redshift is larger than 0.5. 
Furthermore, optical photometry 
(Sect.~\ref{vlt}) allow to estimate that the 
host redshift  is likely lower than 1.7.
When such constraint is used, it is possible to limit the allowed interval for the 
intrinsic $N_{\rm H,z}$ to the range ($0.4\div2.6$)$\times10^{22}$ cm$^{-2}$.

We investigated the possible presence of optical extinction in the host, coupled to 
the observed large X-ray absorption.
First, we computed the spectral energy distribution of the afterglow, taking
advantage of our almost simultaneous VLT $K_{\rm s}$  and XMM observations.
The observed $K_{\rm s}$ band flux falls about two 
orders of magnitude below the extrapolation of the X-ray power law. 
Then, we followed the approach of  \citet{price02} in order to get constraints on 
the expected spectral shape. To calculate the ``Closure'' relations, we assumed the single power law
decay
 described in Sect.\ref{xmm}. The NIR decay is loosely constrained and is compatible 
with the X-ray one. The observed properties of the afterglow are consistent with an isotropic
fireball expansion both into an homogeneous medium, as well as into a stellar wind density
profile environment (see Table~\ref{closure} and its caption for further details). 
\begin{table}
\centering\caption{\label{closure} Constraints on the fireball model. The observed spectral index and
time decay index are used to test different afterglow expansion models: (i) a simple isotropic 
fireball expansion in an homogeneous interstellar
medium \citep[`ISM',][]{sari98}; (ii) an isotropic fireball expansion in a stellar wind density
profile environment \citep[`Wind',][]{chevalier99}; (iii) expansion of a collimated flow 
\citep[`Jet',][]{sari99}. We compute the ``closure'' relation  \citep{price02}
$C = \delta + b\alpha + c$, where $\delta$ is the index of the temporal 
decay and $\alpha$ is the spectral index of the afterglow; the values of the parameters
$b$ and $c$ depend on the model used (ISM, Wind, Jet).
The cases of electron cooling frequency higher ($\nu_c>\nu_X$) or lower ($\nu_c<\nu_X$) than the
EPIC range of sensitivity are considered. A specific model is consistent with the data if the Closure 
value is consistent with 0.}
\label{afterglow}
\begin{tabular}{ccc}
\hline 
Model & Closure ($\nu_c<\nu_X$) & Closure ($\nu_c>\nu_X$)  \\ \hline
ISM  & -0.04$\pm$0.18 & -0.54$\pm$0.18   \\
Wind & -0.04$\pm$0.18 & -1.04$\pm$0.18    \\
Jet & -1.19$\pm$0.22  & -2.19$\pm$0.22   \\ \hline
\end{tabular}
\end{table} 
Both cases require the electron cooling frequency $\nu_c$ to be
lower than the X-ray range (namely, $\nu_c<1.2\times10^{17}$ Hz), a constraint which
may be easily fulfilled at the time of the X-ray observation.
Incidentally, we note that such results agree with 
the ``typical'' X-ray afterglow behaviour 
derived by \citet{piro04} from 36 BeppoSAX follow-up observations of GRBs.
The possible presence of a spectral break between the X-rays and the $K_{\rm s}$ band
(if $\nu_K<\nu_c<\nu_{\rm X}$)
does not allow to conclude that the optical flux is affected by a high extinction; 
of course,  such a possibility may not be ruled out, since $\nu_c$ could be lower than
the NIR range. Both pictures can be easily accomodated within the constraints on the host 
redshift ($0.7<z<1.7$) and gas column density (0.6-2.6$\times10^{22}$ cm$^{-2}$), 
using standard extinction curves \citep{cardelli89}, 
both assuming the Galactic gas-to-dust mass ratio \citep{predehl95} and assuming 
different gas/dust properties in the host \citep[implying an extinction up to 10 times lower,
as suggested by several studies, e.g.][]{galama01,hjorth03,stratta04}.

X-ray absorption in excess wrt. the Galactic one has been already reported
for a handful of GRB afterglows. Thus, to put our results in the right context,
it is useful to review here the results on X-ray absorption in GRB afterglows
(observed a few hours after the burst) obtained so far
by BeppoSAX, Chandra  and XMM-Newton.
\citet{stratta04} present a systematic analysis of a sample of 13
bright afterglows observed with BeppoSAX narrow field instruments (within 5-20 hours
from the prompt);
a significant ($>$99.9\% confidence)
detection of additional N$_{H}$ was found only in two cases (namely,
GRB\,990123 and GRB\,010222), but, owing to the limited photon statistics,
it could not be excluded 
that intrinsic X-ray absorption be 
present also in the other bursts.
Chandra observations of GRB afterglows (starting, on average, $\ga1$ day after the GRB),
did not yield so far useful constraints on the presence of intrinsic X-ray absorption,
apart from a marginal detection for GRB\,020405 \citep[see][and references therein]{stratta04}.

%----------------------------------------------------------- 
   \begin{figure}
   \centering
   \includegraphics[height=8cm,angle=-90]{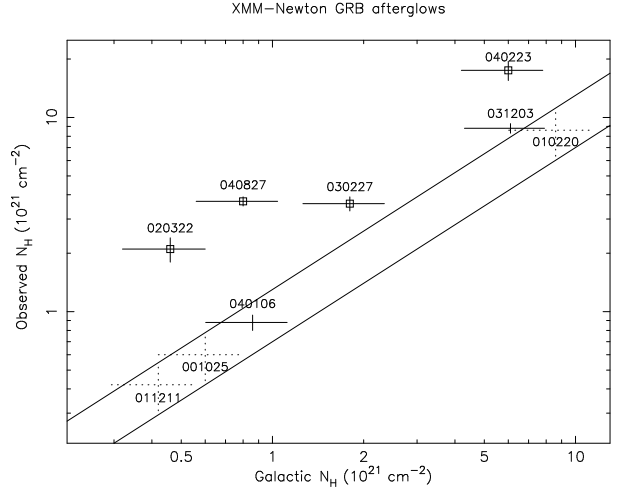}
      \caption{\label{additionalnh} Summary of XMM-Newton results on intrinsic absorption
in GRB afterglows.  
Solid lines correspond to an $N_{\rm H}$ = Galactic $N_{\rm H}$ $\pm$ 30\%,
which represent a reasonable guess \citep[see e.g.][]{stratta04} of the uncertainty affecting Galactic column density estimates.
``Absorbed afterglows'' are marked with a hollow square.
For GRB\,001025, GRB\,010220 and GRB\,011211 the $N_{\rm H}$
value was not evaluated with spectral fitting, but just fixed to the Galactic value,
yielding satisfactory results. We attached to such points an error bar
(in dotted style) corresponding to 30\% of the $N_{\rm H}$.
Reference publications are listed in the 
caption to Table~\ref{nh}.}
   \end{figure}
%
%______________________________________________________________

To date, XMM-Newton has observed 9 GRB afterglows (including GRB\,040827) within a few hours 
($\sim6\div15$) from the burst\footnote{We exclude here the bright afterglow of GRB\,030329, since it could 
be observed only two months after the burst \citep{tiengo03}, owing to satellite pointing constraints.}. 
A detailed summary of XMM-Newton results is reported in Table~\ref{nh} 
as well as in Fig.~\ref{additionalnh}.
Extra absorption has been found in 4 cases, namely GRB\,020322, GRB\,030227, GRB\,040223 and GRB\,040827 (the
best case in such sample - note Fig.~\ref{additionalnh}). 
On the contrary, absorption consistent with the Galactic one was observed 
for GRB\,001025, GRB\,010220, GRB\,011211, GRB\,031203 and GRB\,040106 (the brightest afterglow observed by XMM-Newton so far).

\begin{table*}
\centering\caption{\label{nh} Summary of XMM-Newton observations of intrinsic
gas column density in GRB hosts. Starting from the left,
columns show (1) the GRB Id.; (2) the Galactic $N_{\rm H}$ (cm$^{-2}$) in the GRB direction
according to \citet{dickey90}; (3) the observed $N_{\rm H}$ (cm$^{-2}$), assuming solar system
abundances and a power law model for the afterglow
(for GRB\,001025, GRB\,010220 and GRB\,011211, $N_{\rm H}$ was simply fixed to the Galactic value,
which yielded a satisfactory fit); (4) the allowed range of 
redshifts for the host (`abs': from X-ray spectroscopy using a redshifted
absorber model; `lin': from X-ray spectroscopy, based on emission line detections; 
`opt': from optical spectroscopy); (5) the local gas column density $N_{\rm H,z}$
(cm$^{-2}$) at the
fiducial host redshift reported between parentheses (the value $z=1$
was assumed when no redshift constraints are available). If not 
available in literature, $N_{\rm H,z}$ was computed simply scaling by $(1+z)^{2.6}$ the difference between
observed and expected (Galactic) $N_{\rm H}$; (6) the reference publication: a - \citet{watson02}; b - \citet{mereghetti03};
c - \citet{watson03}; d - \citet{tiengo04}; e - \citet{watson02b}; f - \citet{reeves02}; g - \citet{watson04};
h - \citet{prochaska04}; i - \citet{gendre04}.
}
\begin{tabular}{cccccc}
\hline 
\multicolumn{6}{c}{Absorbed afterglows} \\ \hline
GRB & Galactic $N_{\rm H}$ & Observed $N_{\rm H}$ & z range (method) & $N_{\rm H,z}$ (z)  & Reference
\\ \hline
020322 & 4.6$\times10^{20}$ & (2.1$\pm$0.3)$\times10^{21}$ & 0.7-2.8 (abs) & 1.3$\times10^{22}$ (1.8) & a \\
030227 & 1.8$\times10^{21}$ & (3.6$\pm$0.3)$\times10^{21}$ & 2.5-4.5 (abs) & 6.8$\times10^{22}$ (3.9) & b \\
  ``   & ``  & ``  & 1.33-1.42 (lin) & 3.5$\times10^{22}$ (1.39) & c \\
040223 & 6.0$\times10^{21}$ & (1.8$\pm$0.2)$\times10^{22}$ & ... & 7$\times10^{22}$ (1) & d \\
040827 & 8$\times10^{20}$ & (3.7$\pm$0.2)$\times10^{21}$ & 0.5-1.7 (abs,opt) & 1.0$\times10^{22}$ (1) & this work \\ \hline
\multicolumn{6}{c}{Unabsorbed afterglows} \\ \hline
GRB & Galactic $N_{\rm H}$ & Observed $N_{\rm H}$  & z range (method) & & Reference
\\ \hline
001025 & 6$\times10^{20}$ & = & 0.50-0.55 (lin) &  & e \\
%  ``     & ``  & = & 0-7.1 (abs) & $<$3$\times10^{21}$ (0.5)  & e \\
010220 & 8.6$\times10^{21}$ & =  & 0.97-1.07 (lin) &  & e \\
011211 & 4.2$\times10^{20}$ & = & 2.14 (opt) &  & f \\
031203 & 5.9$\times10^{21}$ & (8.8$\pm$0.5)$\times10^{21}$ & 0.105 (opt) &  & g,h \\
040106 & 8.6$\times10^{20}$ & (8.8$\pm$0.8)$\times10^{20}$ & ... &  & i \\ \hline

\end{tabular}
\end{table*} 

No firm conclusions about the presence of significant, intrinsic
optical extinction may be drawn for the X-ray ``absorbed'' afterglows.
The case of GRB\,020322 \citep{watson02,gendre05} is very similar 
to GRB\,040827.
The lack of an optical afterglow for GRB\,040223 \citep{simoncelli04} could 
also be explained by intrinsic optical extinction within the same hypotheses \citep{gendre05},
although a large foreground extinction is possibly playing an important role for such burst,
owing to its location, close to the Galactic plane.
A complete optical dataset is available only for GRB\,030227, including multicolor, multi-epoch
photometry of the afterglow. Interestingly, in such case \citet{castrotirado03} found ``not feasible''
to connect the X-ray and optical flux distributions within the frame of standard afterglow models, even
considering different extinction laws in the host; therefore, they invoke different emission 
mechanisms at optical and X-ray wavelengths.

In the X-ray ``unabsorbed'' sample, GRB\,011221 
\citep{jakobsson03} and GRB\,040106 \citep{gendre04,moran05} show
very little or no optical extinction.
The case of GRB\,031203 is more problematic:
significant extinction is observed \citep{prochaska04}, although it is 
hard to discriminate the Galactic and host contributions. Moreover,
\citet{malesani04} show that the optical and the X-ray
emission of the afterglow may have a different origin.
No optical emission was detected for the cases of GRB\,001025 \citep[for which the trigger was not promptly 
communicated;][]{smith00} and GRB\,010220 (also lying 
towards the Galactic plane).

Thus, X-ray data provide now conclusive evidence that several GRBs do occur behind a large gas
column density in their host galaxy.
If long GRBs are associated to the death of
massive stars, 
indications of a 
large intrinsic N$_H$ 
absorbing the X-ray afterglows' emission 
would imply that their
progenitors 
are
located within dense 
environments in their host galaxies, most probably star-forming regions
\citep[see e.g.][]{owens98,galama01}. 
In the same scenario, 
the possible
low extinction and reddening 
for the 
optical afterglows
could be explained invoking dust destruction by the early radiation of the GRB
itself,
implying that
the bulk of the absorbing gas 
is
confined
in a compact cloud (10-30 pc) surrounding the GRB progenitor, 
as suggested by previous studies \citep[see e.g.][]{galama01}.
Indeed, the gas column 
densities reported in Table~\ref{nh} are
fully consistent with values observed for inner, overdense regions
of giant molecular clouds in our galaxy 
\citep[see e.g.][and references therein]{lazzati02}, where star formation 
is supposed to take place. The rather shallow constraints on the optical extinction are 
generally consistent with such a picture.

On the other hand, very little may be concluded about the sample of X-ray ``unabsorbed''
afterglows. Lack of statistics in their X-ray spectra coupled to the uncertainty 
on the expected Galactic absorption, together with the uncertainty affecting the host redshift
(based in two cases on highly debated detections of X-ray emission lines) leave room
for at least a few 10$^{21}$ cm$^{-2}$ of gas column density in their host galaxies.
Moreover, such GRBs do not have any
peculiar aspect of their phenomenology clearly distinguishing them wrt.
the ``absorbed'' sample. 
Thus, apparently
there are no reasons to invoke a different
origin for the ``unabsorbed'' GRBs.

Alternatively, the
intrinsic X-ray absorption could be mainly related to the large scale gas distribution
in the GRB host.
In this case, the X-ray phenomenology of the afterglow would
depend on the location of the progenitor
within the galaxy, on
the structure of the host galaxy and on its orientation wrt. the observer,
making it easier to account both for absorbed as well as unabsorbed afterglows.
Under this hypothesis the low optical extinction and reddening could be explained by
different dust-to-gas mass ratios and/or low metallicities in the GRB hosts wrt.
our Galaxy \citep[as suggested by different investigations, e.g.][]{hjorth03,stratta04}.

We note that the problem of inferring properties of the inner
GRB environment from the afterglow's absorption may be further complicated by the 
effects of photo-ionization by the GRB radiation itself \citep{lazzati02};
some evidence for such a process
has been obtained from prompt soft X-ray observations
with BeppoSAX \citep[e.g.][and references therein]{frontera04}.

\section{Conclusions}

Afterglow emission from GRB\,040827 has been studied with XMM-Newton and with
the VLT. 

The observed spectral and temporal properties, when considered in the framework of different
fireball models, are consistent with isotropic expansion into an homogeneous
medium, or into a stellar-wind density profile environment. 
This is a somewhat typical behaviour for X-ray afterglows.

On the other hand, an outstanding peculiarity of GRB\,040827 is the presence of very high intrinsic
X-ray absorption, the best case observed so far, which may be attributed to a gas column density of 
(0.4-2.6)$\times10^{22}~{\rm cm}^{-2}$ in the host galaxy, likely located at $0.5<z<1.7$.
Intrinsic absorption is possibly not an ubiquitous property of GRB X-ray afterglows, 
since it affects $\sim$50\% of the XMM-Newton afterglow observations. 
The presence of high optical extinction accompanying the observed X-ray absorption
cannot be firmly proved, neither for GRB\,040827 nor for other X-ray absorbed afterglows. 

Current X-ray observations do not allow to understand
the nature of the ambient medium surrounding GRB progenitors. 
A major step forward could come only from a study of the evolution of 
the X-ray absorption (coupled to optical extinction) 
in the prompt emission and in the very early phases of the 
afterglow for a relatively
large sample of events. This is now possible with the
Swift satellite \citep{gehrels04}, thanks to its unprecedented capabilities of re-pointing
its X-ray and optical telescopes in a few tens of seconds.

\begin{acknowledgements}
The XMM-Newton data analysis is supported by the Italian Space Agency (ASI).
ADL acknowledges an ASI fellowship.
\end{acknowledgements}

\end{document}

%% file: optlognew.tex
%\documentclass[a4paper,11pt]{aa}
%\begin{document}

\begin{table*}
\caption{GRB040827. Observation log. Observing dates are referred to the middle of the exposures.\label{tb:optlog}}
\begin{tabular}{lllllll}
\hline
\bf Date        &\bf Time since burst  &\bf Filter  &\bf Exposure time &\bf Seeing  &\bf Instrument &\bf Magnitude  \\
(UT)            &(days)                &            &(s)               &(\arcsec)   &                               \\
\hline
2004 Aug 29.99  &2.49                  &$R$         &5$\times$120      &1.2         &VLT+FORS\,1    &$23.55\pm0.50$ \\
2004 Aug 30.04  &2.54                  &$R$         &10$\times$120     &0.6         &VLT+FORS\,2    &$23.02\pm0.15$ \\
\hline
2004 Aug 28.99  &1.50                  &$I$         &4$\times$180      &1.4         &VLT+FORS\,1    &$22.47\pm0.15$ \\
2004 Aug 30.01  &2.52                  &$I$         &5$\times$240      &0.5         &VLT+FORS\,2    &$22.64\pm0.13$ \\
2004 Sep 01.99  &5.51                  &$I$         &10$\times$180     &0.6         &VLT+FORS\,2    &$22.61\pm0.10$ \\
2004 Sep 09.01  &12.51                 &$I$         &6$\times$180      &0.4         &VLT+FORS\,2    &$22.33\pm0.20$ \\
%2004 Sep 16.05 &19.55                 &$I$         &4$\times$240      &0.6         &VLT+FORS\,2    &???            \\
2004 Sep 16.99  &20.50                 &$I$         &8$\times$180      &0.6         &VLT+FORS\,2    &$22.35\pm0.38$ \\
\hline
%2004 Aug 27.98 &0.48                  &$J$         &10$\times$30      &0.9         &VLT+ISAAC      &$20.54\pm0.48$ \\
%2004 Aug 28.02 &0.53                  &$J$         &10$\times$30      &1.0         &VLT+ISAAC      &$21.17\pm0.47$ \\
2004 Aug 28.06  &0.57                  &$J$         &10$\times$30      &1.0         &VLT+ISAAC      &$20.92\pm0.28$ \\
\hline
2004 Aug 27.97  &0.47                  &$H$         &10$\times$30      &0.8         &VLT+ISAAC      &$19.42\pm0.13$ \\
%2004 Aug 28.01 &0.52                  &$H$         &10$\times$30      &0.9         &VLT+ISAAC      &$20.15\pm0.48$ \\
\hline
2004 Aug 27.96  &0.47                  &$K_{\rm s}$ &10$\times$30      &0.7         &VLT+ISAAC      &$19.02\pm0.30$ \\
2004 Aug 28.01  &0.51                  &$K_{\rm s}$ &10$\times$30      &0.9         &VLT+ISAAC      &$18.94\pm0.10$ \\
%2004 Aug 28.05 &0.55                  &$K_{\rm s}$ &10$\times$30      &0.9         &VLT+ISAAC      &$19.80\pm0.57$ \\
2004 Aug 28.98  &1.48                  &$K_{\rm s}$ &60$\times$30      &0.9         &VLT+ISAAC      &$20.05\pm0.13$ \\
2004 Sep 16.02  &19.52                 &$K_{\rm s}$ &30$\times$60      &1.0         &VLT+ISAAC      &$20.08\pm0.15$ \\
2004 Sep 24.01  &27.52                 &$K_{\rm s}$ &45$\times$60      &0.6         &VLT+ISAAC      &$19.88\pm0.10$ \\
2004 Sep 28.00  &31.50                 &$K_{\rm s}$ &48$\times$60      &1.2         &VLT+ISAAC      &$20.35\pm0.20$ \\
\hline\hline
2004 Aug 30.05  &2.54                  &300V        &4$\times$1800     &1.1         &VLT+FORS\,1    &---            \\
\hline
\end{tabular}
\end{table*}

%\end{document}